\documentclass[twocolumn,amsmath,amssymb,pra,longbibliography]{revtex4-1}

\usepackage{times}
\usepackage{graphicx}
\usepackage{amssymb}
\usepackage{url}
\usepackage{epstopdf}
\usepackage[unicode]{hyperref}
\hypersetup{
   unicode=true,          
   plainpages=false,
   colorlinks=true,       
   citecolor=blue,
}
\newcommand{\vs}{v_{\mathrm{s}}}
\newcommand{\vc}{v_{\mathrm{c}}}
\newcommand{\mI}{m_{\mathrm{I}}}
\newcommand{\Es}{E_{\mathrm{s}}}

\newcommand{\Nd}{N_{\mathrm{d}}}
\newcommand{\NdDS}{N_{\mathrm{d}}^\mathrm{GP}}
\newcommand{\Ps}{P_{\mathrm{s}}}

\newcommand{\pf}{p_{\mathrm{F}}}
\newcommand{\vcf}{v_{\mathrm{cf}}}

\newcommand{\sfs}{{\sigma_\mathrm{fs}}}
\newcommand{\sDS}{{\sigma_\mathrm{GP}}}
\newcommand{\nbg}{{n_\mathrm{bg}}}

\newcommand{\Dx}{\Delta X}
\newcommand{\Xo}{X_0}
\newcommand{\X}{X}

\begin{document}
\title{Quantum dark solitons in the one-dimensional Bose gas}
\author{Sophie S. Shamailov}
\email{s.shamailov@auckland.ac.nz}
\altaffiliation{Present address: Dodd-Walls Centre for Photonics and Quantum Technology, Department of Physics, University of Auckland, Private Bag 92019, Auckland, New Zealand}
\affiliation{Dodd-Walls Centre for Photonics and Quantum Technology, New Zealand}
\affiliation{New Zealand Institute for Advanced Study,  Centre for Theoretical Chemistry and Physics, Massey University, Private Bag 102904, North Shore, Auckland 0745, New Zealand}
\author{Joachim Brand}
\email{j.brand@massey.ac.nz}
\affiliation{Dodd-Walls Centre for Photonics and Quantum Technology, New Zealand}
\affiliation{New Zealand Institute for Advanced Study,  Centre for Theoretical Chemistry and Physics, Massey University, Private Bag 102904, North Shore, Auckland 0745, New Zealand}
\pacs{02.70.-c, 03.75.Lm, 03.65.-w, 05.60.Gg}
\keywords{dark solitons, repulsive interactions, nonequilibrium quantum dynamics}
\date{\today} 

\begin{abstract}
Dark and grey soliton-like states are shown to emerge from numerically constructed superpositions of translationally-invariant eigenstates of the interacting Bose gas in a toroidal trap. The exact quantum many-body dynamics reveals a density depression with ballistic spreading that is absent in classical solitons. A simple theory  based on finite-size bound states of holes with quantum-mechanical center-of-mass motion quantitatively explains the time-evolution and predicts quantum effects that could be observed in ultra-cold gas experiments. The soliton phase step is found relevant for explaining finite size effects in numerical simulations. 
An invariant fundamental soliton width is shown to deviate from the Gross-Pitaevskii predictions in the interacting regime and vanishes in the Tonks-Girardeau limit.
\end{abstract}

\maketitle

\section{Introduction}
Dark solitons \cite{Tsuzuki1971}  are ubiquitous features of superfluids and have been observed frequently in ultra-cold atomic gas experiments 
\cite{Denschlag2000,Burger1999,ginsberg05,Becker2008,Weller2008,Lamporesi2013,Ku2016,Yefsah2013}. 
The characteristic localised density depression is stabilised by the competing effects of hydrostatic pressure and the stiffness of the superfluid phase. While experiments to date could be well explained by classical theory,
there has been much debate about quantum effects
\cite{Yefsah2013,Astrakharchik2012,Dziarmaga2004,Mishmash2009}. Quantum features of dark solitons are expected to be particularly relevant under reduced dimensionality,
where quantum fluctuations 
destroy long-range coherence of the superfluid phase. While theoretical works  on the one-dimensional Bose gas have predicted effects like greying of the dark soliton \cite{Dziarmaga2002,Dziarmaga2002a,Dziarmaga2004,Mishmash2009,Delande2014},  and have pointed to a connection of dark solitons to quantum-many-body eigenstates of the Bethe-ansatz solvable Lieb-Liniger model \cite{Kulish1976,Ishikawa1980,Astrakharchik2012,Kanamoto2010,Fialko2012b,Roussou2017,Jackson2011,Syrwid2015,Sato2012a,Sato2016}, the full picture connecting the physical effects  with the exact eigenstates is still missing. 

Specifically, Ref.\ \cite{Ishikawa1980} showed that the dispersion relation of yrast states (eigenstates with lowest energy at given momentum) in the Lieb-Liniger model asymptotically approaches that of dark solitons in the  Gross-Pitaveskii (GP)  or classical nonlinear Schr\"odinger equation in the high-density limit. However, in contrast to the translationally-invariant yrast states of constant particle density, classical dark solitons have a localised density dip that propagates with constant velocity. On the other hand, numerical simulations of single-shot measurements of particle position in the yrast states show localised voids appearing at random positions \cite{Syrwid2015,Syrwid2016}. Superpositions of yrast states were further shown to exhibit translational symmetry breaking  under weak interactions \cite{Fialko2012b,Roussou2017}, and localised density depressions at finite interactions that decay during time evolution \cite{Sato2012a,Sato2016}. 
However, control over soliton parameters, the classical limit, or quantitative understanding of beyond mean-field effects were not achieved. 

The situation is  better understood for bright solitons, where quantum effects were observed in optics experiments \cite{Rosenbluh1991,Drummond1993} and a full quantum theory was developed by constructing quantum soliton states as superpositions of translationally invariant eigenstates of an interacting boson model \cite{McGuire1964,Lai1989,Cosme2015a,Ayet2017}.

In this work we bridge the gap in the quantum theory of dark solitons by constructing quantum many-body states that most-closely resemble classical dark solitons from superpositions of yrast eigenstates, and quantifying their properties. We simulate the full quantum dynamics making use of exact solutions from the Bethe ansatz. While the behavior of classical dark solitons is recovered in the high density limit, we observe ballistic spreading in the crossover to the low-density, strongly-correlated limit, known as the Tonks-Girardeau gas. Modeling the quantum dark soliton as a finite-size quantum mechanical quasiparticle (inspired by Ref.~\cite{Konotop2004}), we identify the velocity, a soliton mass, and a fundamental soliton width as characteristic parameters for the dynamics of the  simulated density depletion. These parameters can be obtained from the yrast dispersion relation with finite size corrections, attaining excellent  agreement with the numerical simulations. The particle number depletion and a quantity interpreted as the soliton phase step play important roles in the finite size corrections and can also be computed from the dispersion relation.

\section{Yrast states in the Lieb-Liniger model} 

We model a gas of  $N$  bosonic atoms with mass $m$ in a tightly-confining toroidal trap of circumference $L$ by the Lieb-Liniger model \cite{lieb63:1,lieb63:2} with repulsive  interactions $c>0$ \footnote{Note that $c = -2/a_{1D} = g m/\hbar^2$, where $a_{1D}$ is the 1D scattering length  and $g$ the effective coupling constant
\cite{olshanii98}.}
\begin{align}
\label{Ham}
\hat{H} = -\frac{\hbar^2}{2m} \sum_{i=1}^N\frac{\partial^2}{\partial x_i^2} + \frac{\hbar^2 c}{m}\sum_{i<j} \delta(x_i-x_j) .
\end{align}
The eigenstates of $H$ can be constructed with the Bethe ansatz  from the set of $N$  rapidities $\{k_j\}$, which in turn is fully determined by $N$ quantum numbers $I_j$ through the Bethe equations
\begin{align} \label{eq:Bethe}
k_j+\frac{1}{L}\sum_l 2 \arctan\frac{k_j-k_l}{c}=\frac{2 \pi}{L}I_j ,
\end{align}
where the $I_j (+\frac{1}{2})$ take integer values for odd (even) $N$ \cite{Yang1969}. While the momentum  $P=\hbar\sum_j k_j = 2\pi\hbar/L\sum_j I_j$ is already determined by the quantum numbers $I_j$, the energy $E=\hbar^2/{2m}\sum_j k_j^2$ depends on the interaction strength through the rapidities. Of particular relevance are \emph{yrast} states denoted by $|P,\mathrm{yr}\rangle$, which are the eigenstates of lowest energy $E_P^N$ for given $P$ and $N$. They are found from otherwise contiguous sets of $I_j$ with a gap of up to one quantum number.

\section{Quantum dark solitons} 
We construct initial states as Gaussian superpositions of yrast eigenstates centered around $P_0$ with width $\Delta P$:
\begin{align} \label{eq:qds}
|{P_0}\rangle =& \sum_q C_q^{P_0}  |q,\mathrm{yr}\rangle,
\\ C_q^{P_0} =& A e^{-\frac{(q-P_0)^2}{4\Delta P^2} + i \frac{q\Xo}{\hbar}} , \label{eq:Gauss}
\end{align}
where 
$\Xo$ is  a displacement. The time evolution is given by $|{P_0}(t)\rangle = \exp(-i\hat{H}t/\hbar)|{P_0}\rangle$. 
As the main observable, we construct the single-particle density 
$n(x,t)= \langle{P_0}(t)| \hat{\rho}(x) |{P_0}(t)\rangle$ as
\begin{align} \label{eq:densityevolution}
\nonumber
n(x,t) 
= & \sum_{p,q}C_q^{P_0*}C_p^{P_0} \langle{q,\mathrm{yr}}| \hat{\rho}(0) |{p,\mathrm{yr}}\rangle\\
& \times \exp[i(p-q)x/\hbar - i(E_p-E_q)t/\hbar] ,
\end{align}
where the density form factor $\langle{q,\mathrm{yr}}| \hat{\rho}(0) |{p,\mathrm{yr}}\rangle$
is calculated from the rapidities $\{k_j\}$ 
using formulas derived from the algebraic Bethe ansatz \cite{1989_Slavnov_TMP_79,1990_Slavnov_TMP_82,1982_Korepin_CMP_86,Caux2007,Sato2012a}. 
Density profiles of equal-weight superpositions over all yrast states were previously shown to produce localised but rapidly dispersing depressions translating at different velocities  from those of  fitted GP dark soliton profiles \cite{Sato2012a,Sato2016}.

\begin{figure}[ht!]
\includegraphics[width=0.95\columnwidth]{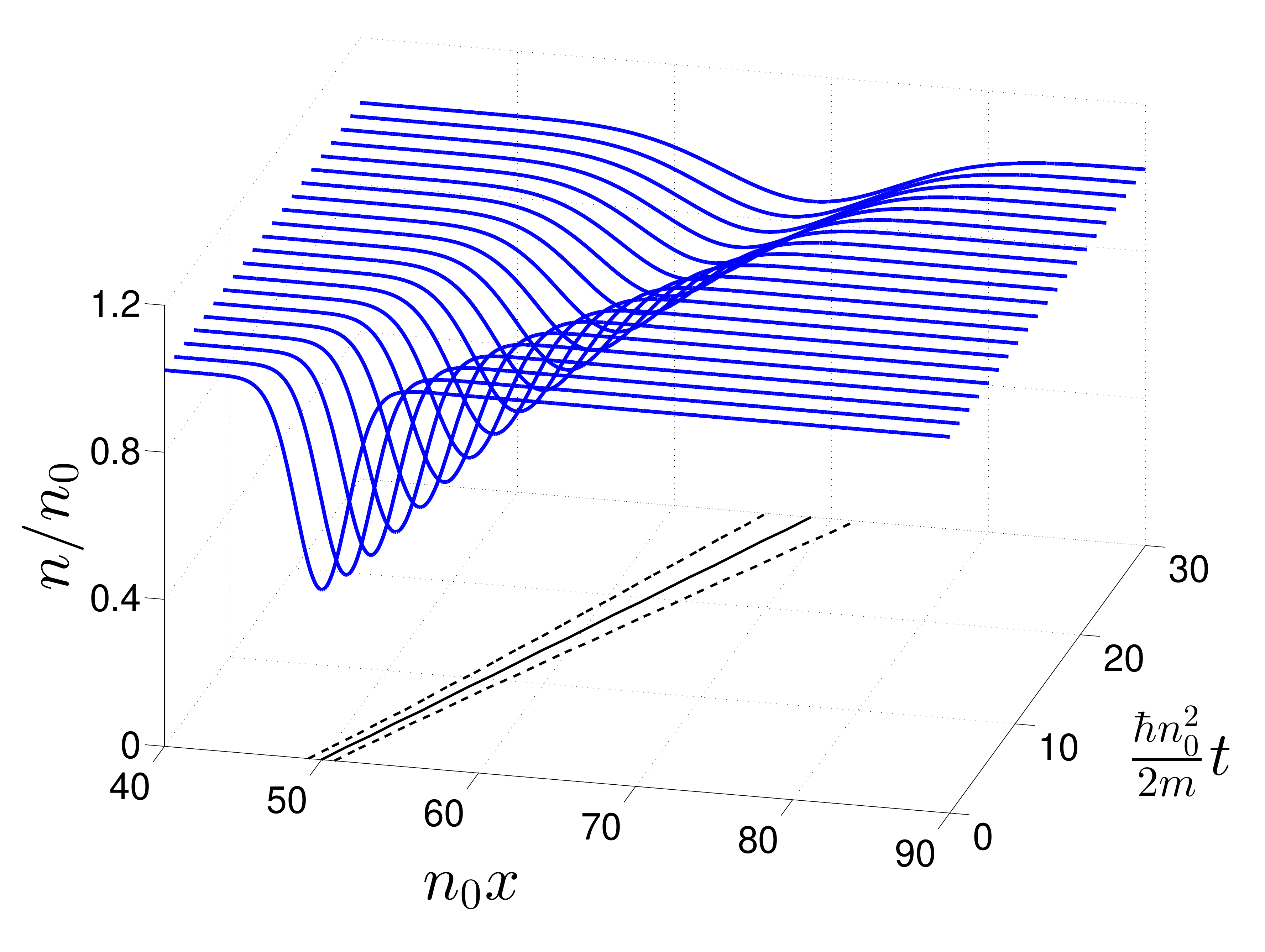}\\
\caption{Time evolution of the quantum dark soliton \eqref{eq:qds} constructed as a superposition of yrast eigenstates of the Lieb-Liniger model with $N=100$ particles at the intermediate interaction strength $\gamma=1$, where $\gamma = c /n_0$ and $n_0= N/L$. The superposition is prepared with $\Delta P = 0.11\pi\hbar n_0$ and $P_0 = 0.64\pi\hbar n_0$. The solid line tracks the minimum of the dip and the dashed lines on either side of it are displaced by half  of the soliton's width, i.e.~by $\pm \Dx/2$ [see Eq.~(\ref{eq:width})].
}\label{fig:density}
\end{figure}

Figure \ref{fig:density} shows the time evolution of the density profile with initial state \eqref{eq:qds}. Numerical simulations with varying parameters consistently show a smooth and localised density dip that propagates at constant velocity $\vs$ with $\X(t) =  \Xo - \vs t$ while the width $\Dx$ increases over time. Here, $\X\equiv\overline{x}$ measures the position, and the variance 
\begin{align}
\label{eq:width}
\Dx^2 =\overline{x^2} - \overline{x}^2
\end{align}
the width. The average $\overline{A} = \int A \tilde{n}\ dx / N_\mathrm{d}$ is evaluated with respect to the density deviation $\tilde{n} = n(x) - n_\mathrm{bg}$ from the constant background $n_\mathrm{bg}$, 
where $N_\mathrm{d}= \int \tilde{n}\ dx$ is the particle number depletion. In our time-dependent simulations, $N_\mathrm{d}$ remains approximately constant over time. Motion at constant velocity and $N_\mathrm{d}$  with expanding width (i.e.\ ``greying of the dark soliton'') are exactly as expected for quantum dark solitons \cite{Dziarmaga2002a,Law2003,Mishmash2009a,Dziarmaga2004,Dziarmaga2002}. 

\section{Theory of quantum dark solitons} 
We aim to formulate a quantitative theory of the observed propagation at constant velocity $\vs$ and the spreading of the soliton width. In analogy to the case of bright quantum solitons \cite{Cosme2015a,Lai1989}, which consist of finite-size bound states of bosons with a quantum mechanical center-of-mass motion, we assume that the variance of the solitonic dip in the single-particle density of a Gaussian superposition state, $\Dx^2$, can be decomposed as
\begin{align} \label{eq:varsum}
\Dx^2(t) = \sfs^2+ \sigma_\mathrm{CoM}^2(t),
\end{align}
where $\sfs^2$ is the variance of the fundamental soliton, which is constant in time and independent of the  superposition parameters $\Delta P$ and $\Xo$ \footnote{A corresponding result to Eq.\ \eqref{eq:varsum} was proved in Ref.\ \cite{Cosme2015a}}. 
The center-of-mass variance $\sigma_\mathrm{CoM}^2(t)$ follows the time evolution of a Gaussian wave-packet in the single-particle Schr\"odinger equation, given by
\begin{align} \label{eq:comvar}
\sigma_\mathrm{CoM}^2(t) = \sigma_0^2\left[1+\left(\frac{\hbar t}{2 M \sigma_0^2}\right)^2\right] ,
\end{align}
where
\begin{align} \label{eq:initvar}
\sigma_0^2 = \frac{\hbar^2}{4 \Delta P^2}
\end{align}
is the initial variance of the Gaussian wave-packet density in real space and $M$ is a mass parameter. The quadratic-in-time growth of the variance 
is characteristic of  ballistic motion and is faster than diffusion \footnote{In regular diffusion the growth of the variance is linear in time. The term ``quantum diffusion'' for quantum solitons that is found in the literature \cite{Dziarmaga2004} is thus a misnomer.}. The same effect is expected for bright solitons \cite{Lai1989,Cosme2015a,Ayet2017}. 

\begin{figure}[ht!]
\includegraphics[width=\columnwidth]{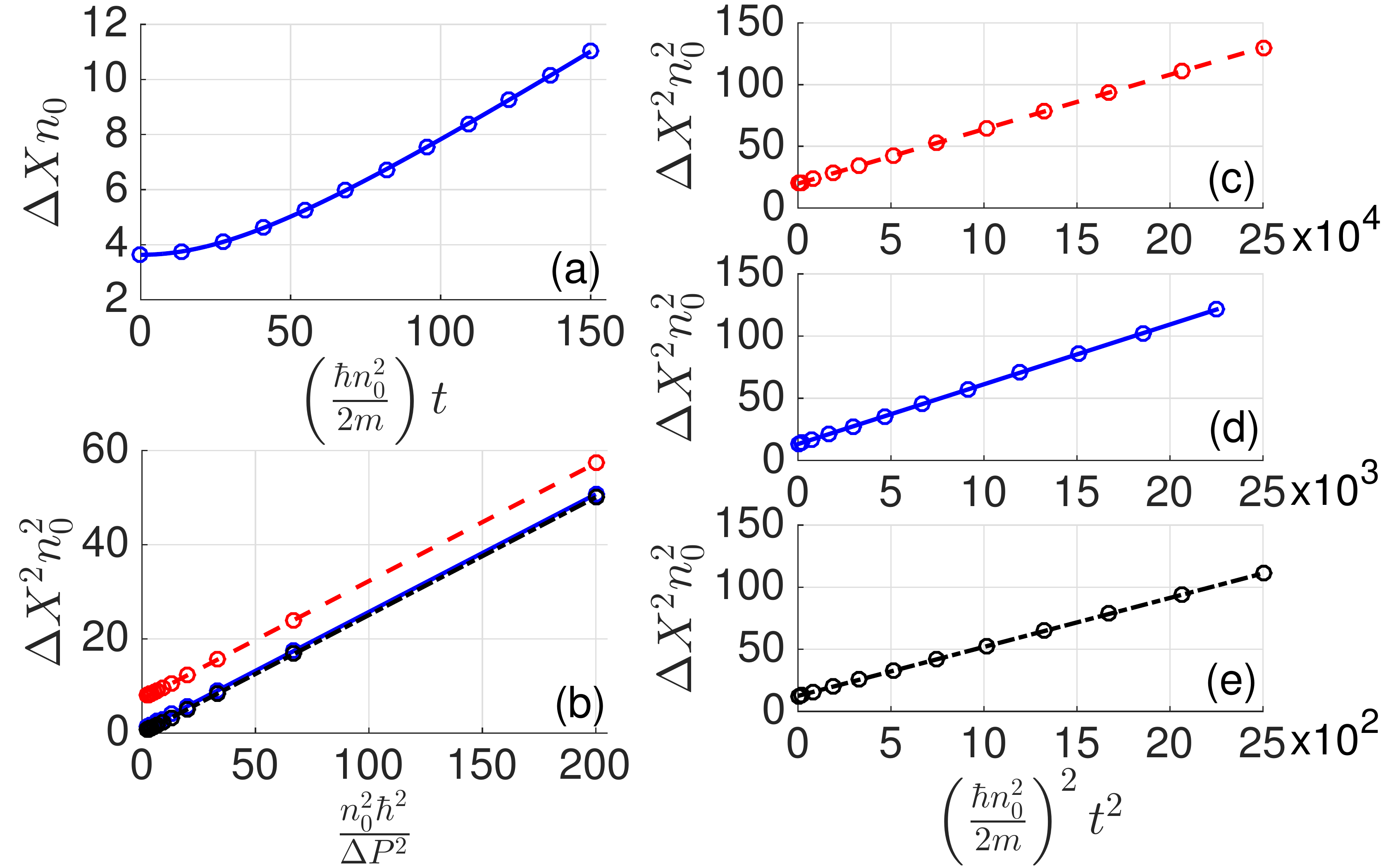}\\
\caption{Width $\Dx$ of the density depression from simulations (symbols) compared with fits of Eq.~\eqref{eq:varsum} (lines). (a) Ballistic growth of $\Dx$ in time for  $\gamma=1$ with $\sfs$ and $M$ fitted. (b) Initial variance $\Dx^2$ at $t=0$ vs.\ $\Delta P^{-2}$ used for extracting $\sfs^2$ as the intercept. Interaction strengths $\gamma = 0.1, 1,10$ are shown by red dashed, blue continuous and black dotted lines, respectively. Panels (c), (d), (e) show linear fits of $\Dx^2$ vs.\ $t^2$ used to extract $M$ for  $\gamma = 0.1, 1,10$, respectively, with $\Delta P = 0.045\pi \hbar n_0$. All panels used $P_0=\pi\hbar n_0$ and  $N=100$. The expected quadratic dependence of the variance on $1/\Delta P$ and $t$ is evident in all parameter regimes.
}\label{fig:phenofit}
\end{figure}

The three constant parameters -- the soliton velocity $\vs$, fundamental width $\sfs$, and mass $M$ -- completely characterise the motion of the first and second moment of the quantum dark soliton according to Eqs.\ \eqref{eq:varsum} -- \eqref{eq:initvar}. We have performed extensive quantum simulations of the density profile with Eq.\ \eqref{eq:densityevolution} and found excellent agreement with this model for a wide range of parameters, as shown in Fig.\ \ref{fig:phenofit} (as long as $L \gg \Dx$ and $\Delta P \ll \pi \hbar n_0$). Interpreting quantum dark solitons as quasi-particles in Landau's sense \cite{Konotop2004}, it is not surprising that the soliton velocity observed in simulations agrees with the group velocity $dE/dP$ and the mass parameter $M$ with the inertial mass $(d^2E/dP^2)^{-1}$ of the yrast dispersion relation. 

\section{Yrast dispersion relation} 
The yrast excitation energy $E_P^N-E_0^N$ becomes a continuous function  $\Es^\infty(P)$ of momentum in the thermodynamic limit where $N,L\to\infty$ while $n_0=N/L$ remains constant. The continuous dispersion relation can be obtained by solving Fredholm integral equations \cite{lieb63:2} and is useful for obtaining various relevant properties for the quasiparticle description as derivatives, e.g.\ the quasiparticle velocity $\vs=d\Es^\infty/dP$ and inertial mass $\mI^{-1} = d^2\Es^\infty/dP^2$, pertaining to an infinite system. In order to obtain quantitative agreement with our numerical simulations, finite-size corrections need to be applied. The leading $1/L$ correction terms
is found 
from a conceptually-simple argument 
assuming
that yrast states are associated with (soliton-like) quasiparticles 
with two features, in particular:
(a) A particle number depletion $\Nd$ arising from a density dip that is localised on a scale that is small compared to the box size $L$, which leads to an elevated background density $\nbg= n_0-\Nd/L>n_0$, and (b) a nominal ``phase step'' $\Delta \phi$ that leads to a backflow current with velocity $\vcf=\hbar\Delta \phi/mL$. This background current corresponds to a linear phase gradient that connects the phase step at the soliton across the periodic boundary conditions.
The soliton moving on the background experiences a Galilean boost. The  finite system dispersion relation to leading order $\mathcal{O}(L^{-1})$ is then obtained from
\begin{eqnarray} 
&& E_P^N-E_0^N \approx \Es^N(P) \equiv \nonumber\\
&& \Es^\infty(P) + P_s \vcf + \frac{1}{2}N m \vcf^2 - \frac{\Nd^2}{2L}\frac{d\mu}{dn_0},
\label{eq:finiteDispersion}
\end{eqnarray}
where $P_s = \Nd m \vs$ is the physical momentum of the moving density depletion and the last term is a correction of the ground state energy due to the localised particle depletion obtained from a Taylor expansion of the equation of state. All quantities on the right hand side of Eq.~\eqref{eq:finiteDispersion} are evaluated in the thermodynamic limit at the  background density $n_\mathrm{bg}$.

\begin{figure}[ht!]
\includegraphics[width=\columnwidth]{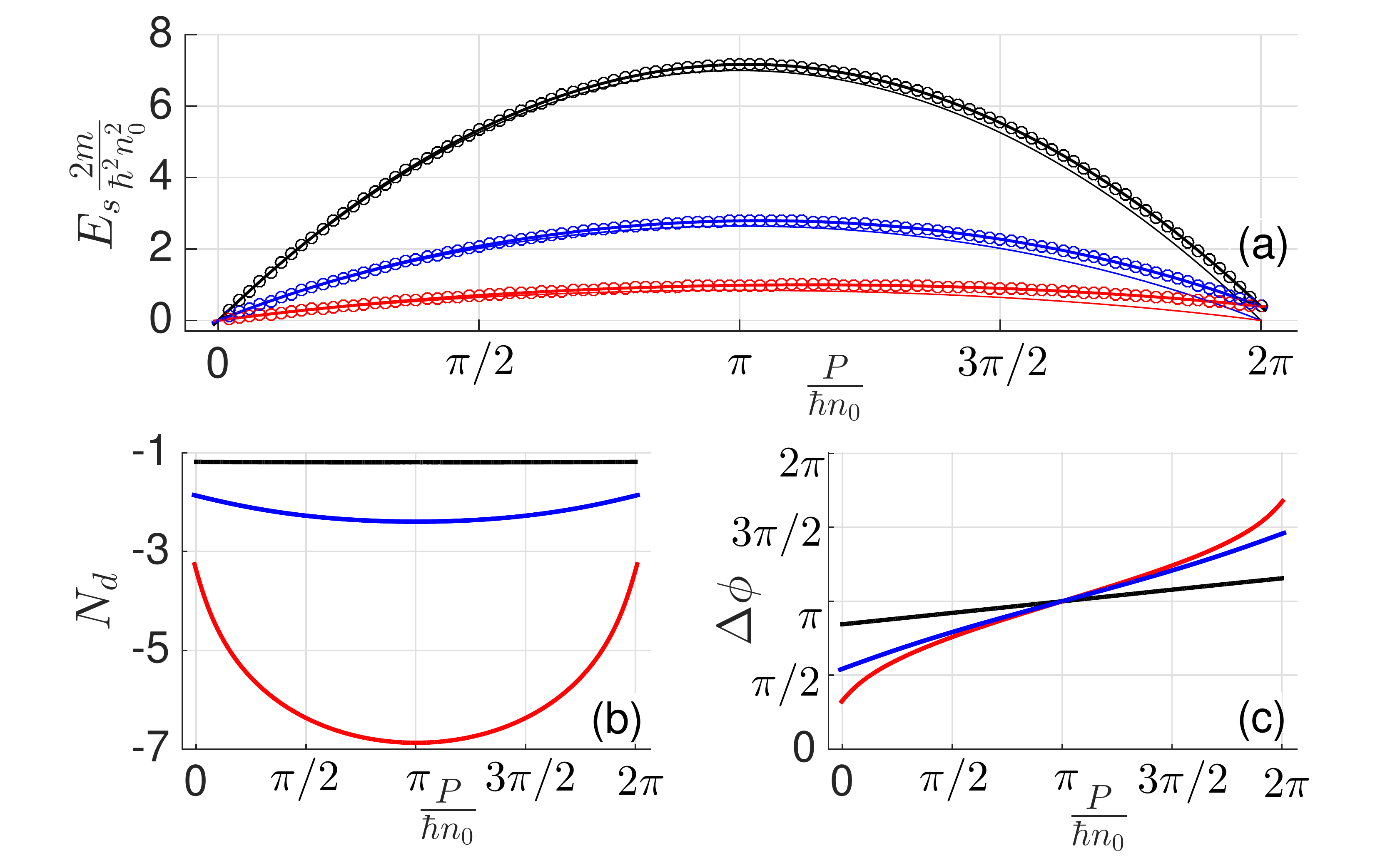}\\
\caption{(a) Dispersion relation of yrast states of the Lieb-Liniger model. Symbols show the excitation energies  $E_P^N-E_0^N$ for $N=100$ vs.\ momentum. Thick lines show the approximate formulae for the finite system \eqref{eq:finiteDispersion} and thin lines show the dispersion relations $\Es^\infty(P)$ in the thermodynamic limit \cite{lieb63:2} for comparison. The interaction strengths are $\gamma=0.1$ (dashed red line \& red circles), $\gamma=1$ (full blue line \& blue circles), and $\gamma = 10$ (dash-dotted black line \& black circles); the same colour code is used in the bottom panels.  Bottom panels: particle number depletion (c) and phase step (d) with finite size corrections for $N=100$.
}\label{fig:dispersion}
\end{figure}

The Galilean boost demands that $P=\Ps + N m \vcf$, which can be used to determine the backflow velocity $\vcf$, and hence the phase step $\Delta \phi$, once the particle number depletion $\Nd$ is known. The latter can be computed from the dispersion relation as \cite{Schecter2012,Shamailov2016}
\begin{align} \label{eq:Nd}
\Nd = - \left(1-\frac{\vs^2}{\vc^2} \right)^{-1} \left(\frac{\partial \Es^{\infty}}{\partial \mu} + \frac{\vs P}{m\vc^2}  \right) ,
\end{align}
where the derivative has to be taken at constant $P$ and $c$, $\vc$ is the speed of sound defined by $m\vc^2 = n_0 \,d\mu/dn_0$, and $\mu = \lim\limits_{N\to\infty} d E_0^N / dN$ is the chemical potential of the ground state. Equation \eqref{eq:Nd} was derived under similar assumptions to (a) and (b).
For GP dark solitons in an infinite box the assumptions hold and Eq.~\eqref{eq:Nd} becomes exact. The dispersion relation is show in Fig.~\ref{fig:dispersion} (a). Both $\Nd$ and $\Delta\phi$ are shown in the bottom panels of Fig.~\ref{fig:dispersion}. Finite size corrections to these quantities simply amount to solving the thermodynamic limit Bethe ansatz equations and evaluating $\Nd$ and $\Delta\phi$ at the elevated background density $n_{\mathrm{bg}}$.

Even though the assumptions of a localised density dip (a) and a phase step responsible for a superfluid current (b) are not obviously satisfied for type-II Lieb-Liniger states, we find that, as for GP dark solitons, the continuous approximation of the dispersion relation is excellent in all interaction regimes as long as $\sfs \ll L$ \footnote{This condition is easily violated near the edges of the dispersion relation, i.e.~$P_0\approx 0, 2\pi\hbar n_0$, and for very weak interactions in a finite box.} (see Fig.~\ref{fig:dispersion}). In the Tonks-Girardeau limit of $\gamma\rightarrow\infty$ the approximation \eqref{eq:finiteDispersion} becomes exact with $\Nd=-1$, $\Delta\phi=\pi$ and $\Es^N(P) = [-P^2 +2P\pf(1+N^{-1})]/2m$, where $\pf=\pi n_0 \hbar$ is the Fermi momentum. 
This approximation works very well in all regimes, which implies that  the concepts of a phase step and global backflow current are useful 
despite the fact that global phase coherence is not expected due to strong fluctuations in 1D leading to algebraic off-diagonal long-range order.

\begin{figure}[ht!]
\includegraphics[width=\columnwidth]{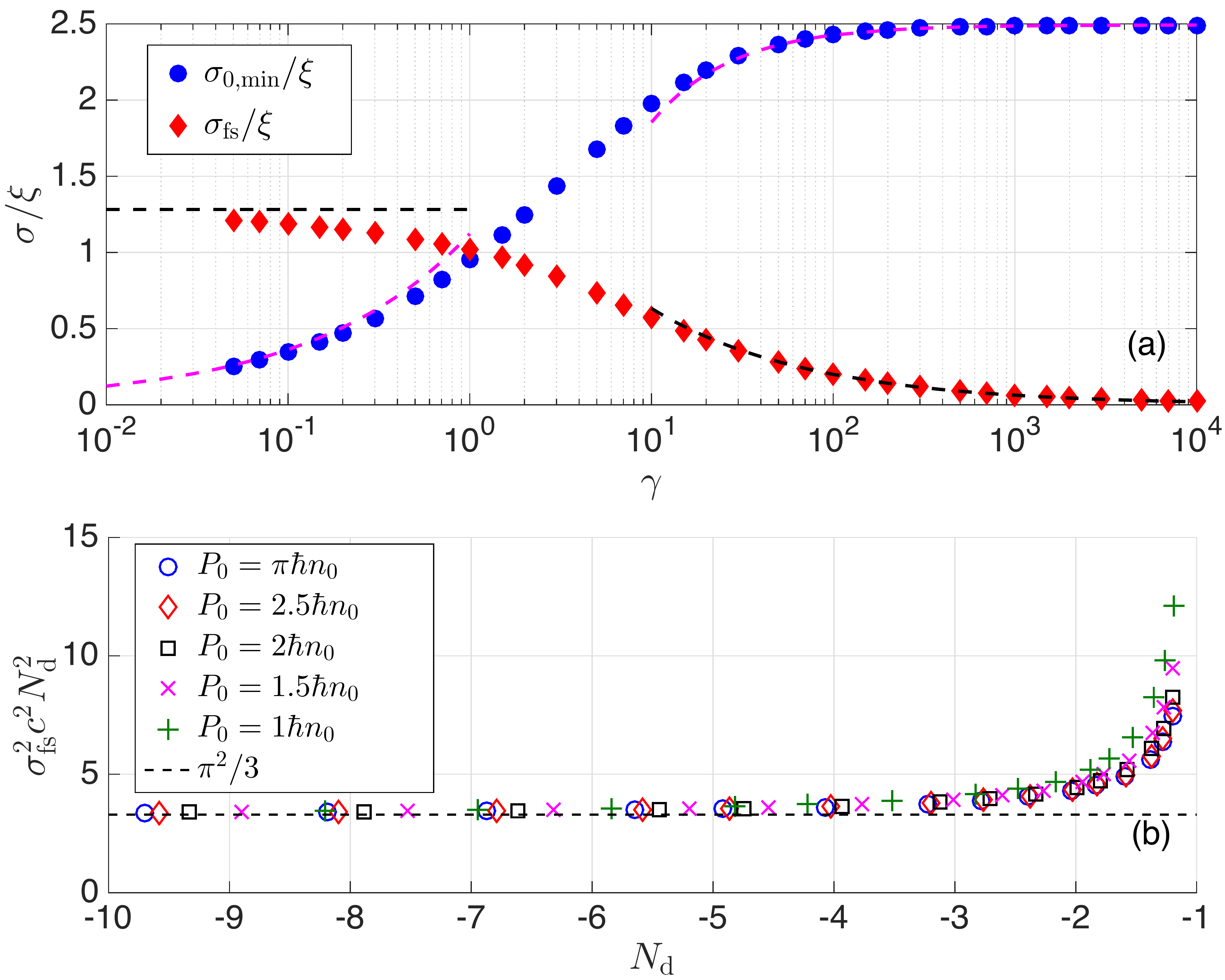}\\
\caption{Length scales of the quantum dark soliton. (a) Fundamental soliton width $\sfs$ and minimum center-of-mass wave-packet width $\sigma_{0,\textrm{min}}\approx 0.8 n_0^{-1}$ vs.\  coupling strength $\gamma = c /n_0$ for $P_0 = \pi\hbar n_0$. Limiting analytical approximations: $\sfs/\xi \to \pi/\sqrt{6}$ from GP theory for $\gamma \ll 1$ and Eq.~\eqref{eq:sfslargegamma} for $\gamma \gg 1$ (black dashed lines). Magenta dashed lines:  $\sigma_{0,\textrm{min}}/\xi$ with $\xi \sim 1/(n_0 \sqrt{2\gamma})$ for $\gamma \ll 1$ and $\xi \sim n_0^{-1}\pi^{-1}(1+8/3\gamma)$ for $\gamma \gg 1$. (b) Numerical data for $\sfs^2$ (multiplied by $c^2 \Nd^2$) vs.~the particle number depletion $\Nd$ from Eq.~\eqref{eq:Nd} (with finite size corrections). Data from different momenta and  coupling strengths collapse onto the same curve and deviate from the result $\sigma_{\mathrm{GP}}^2$ (dashed line) for the classical dark soliton only near the Tonks-Girardeau limit where $\Nd\to -1$. The width $\sfs$ was extracted from simulation data with $N=100$.
}\label{fig:fundamental}
\end{figure}

\section{Length scales} 
In contrast to a classical soliton, which propagates with constant shape, the density profile of the quantum dark soliton changes in time. According to Eqs.~\eqref{eq:varsum} \& \eqref{eq:comvar} the strongest localization occurs  at $t=0$ and is determined by the fundamental soliton width $\sfs$ together with the length scale of the Gaussian wave packet $\sigma_0$. The choice of the latter is 
limited by the requirement of $\Delta P$ fitting in to the fundamental momentum interval $[0,2\pi\hbar n_0]$.
We estimate the minimal value
$\sigma_{0,\textrm{min}}$ conservatively from Eq.~(\ref{eq:initvar}) with $\Delta P \lesssim \pi\hbar n_0/5$. Figure \ref{fig:fundamental} (a) shows the two length scales $\sfs$ and $\sigma_{\mathrm{0,min}}$ crossing over at intermediate interactions, with the size of the quantum dark soliton limited by the larger length scale.

The fundamental soliton width $\sfs$ is an interesting nontrivial quantity that we extract from numerical simulations by fitting [see Fig.~\ref{fig:phenofit} (b)]. For small $\gamma$ our data agree very well with the dark soliton width computed from the GP equation according to Eq.~\eqref{eq:width}, $\sDS=\pi \xi /\sqrt{6(1-\vs^2/\vc^2)}$, where $\xi=\hbar/\sqrt{2m\mu}$, while for large $\gamma$ the fundamental soliton width $\sfs/\xi$ tends to zero. Close inspection reveals that 
\begin{equation} \label{eq:sfslargegamma}
\sfs/\xi \approx 2/\sqrt{\gamma} \quad\textrm{for} \quad \gamma \gg 1
\end{equation}
fits the numerical data very well [see Fig.~\ref{fig:fundamental} (a)]. 
The vanishing of $\sfs$ demonstrates that the fundamental soliton changes from a macroscopic object in the Bogoliubov regime, where it coincides with the GP dark soliton, to a single-particle hole without an intrinsic length scale in the Tonks-Girardeau limit. 

It is tempting to interpret the quantum dark soliton as a bound state of  $|\Nd|$ holes (a fractional number) in analogy to quantum bright solitons, which are bound states of $N$ bosons \cite{Lai1989}, where the fundamental soliton width is a length scale of the multi-particle bound state \cite{Cosme2015a}. Indeed, the length scale $\sDS$ for the GP dark soliton can be re-expressed as $\sDS = \pi/(\sqrt{3} c |\NdDS|)$, where the velocity-dependence is fully subsumed in the particle number depletion $\NdDS$. Plotting numerical data for $\sfs$ vs.\ $\Nd$ in Fig.~\ref{fig:fundamental} demonstrates that data taken at different interaction strengths $\gamma=c /n_0$ and momenta $P_0$ falls onto a single curve within numerical accuracy, which means that $\sfs$ also appears to depend directly only on $\Nd$ and $c$. Significant deviations from the GP formula are observed only close to $\Nd = -1$, which corresponds to the strongly correlated Tonks-Girardeau limit. 

Interpreting the quantum soliton as a bound state of holes with quantum-mechanical center-of-mass motion is consistent with lattice simulations at small $\gamma$ \cite{Delande2014}. These showed that imprinted dark solitons display an innate soliton profile with constant length scale in single-shot images, while the single-particle density displays a spreading and weakening depression over time due to a growing uncertainty over the soliton position.  Our results quantify these effects and suggest that the same physical picture is relevant far into the strongly correlated regime.

Classical solitons emerge in our theory in the Bogoliubov limit $\gamma\to0$, where $\sfs\to\pi\xi/\sqrt{6}=\pi/\sqrt{12\gamma}n_0$ and $M\to2m\Nd\to -4m\sqrt{1-\vs^2/\vc^2}/\sqrt{\gamma}$ become macroscopic. Constructing a wave packet with $\epsilon \equiv \Delta P/ 2\pi n_0\hbar\ll 1$, 
we find that the initial soliton can be well localised ($\sigma_0\ll\sfs$) when $\epsilon^2\gg 3 \gamma/4\pi^4$ and remains so ($\sigma_\mathrm{CoM}^2-\sigma_0^2\ll\sfs^2$) for a time $t\ll\sqrt{1-\vs^2/\vc^2} m/(\sqrt{6}\gamma\epsilon\hbar n_0^2)$. We have further verified that  numerical density profiles at $\gamma=0.01$ are nearly indistinguishable from GP solitons at the same momentum $P_0$.

\section{Conclusions} 
The  yrast states of the Lieb-Liniger model are strongly correlated, fragmented \cite{Fialko2012b,Odziejewski2018}, and contain relevant information about the solitonic dip in high order correlation functions \cite{Syrwid2015}. In this situation it may seem remarkable and surprising that solitonic physics can be extracted from the single-particle density of superposition states and easily quantified by the  hypothesized equations (\ref{eq:varsum}) -- (\ref{eq:initvar}). 
On the other hand it is known from the theory of quantum bright solitons, that wave-packet superpositions of fragmented and translationally invariant eigenstates can achieve almost unit condensate fraction \cite{Ayet2017}. Such states are only weakly correlated and closely resemble bright solitons of typical ultra-cold gas experiments (e.g.~Ref.\ \cite{Khaykovich2002}). While our computational approach does not provide access to the condensate fraction, there is nevertheless good reason to believe that the initial superposition states of our simulations [Eq.\ \eqref{eq:qds}] for small $\gamma$ are weakly correlated as well and closely resemble the quantum states prepared in dark soliton experiments with Bose-Einstein condensates, e.g.~in Refs.~\cite{Burger1999,Becker2008,Weller2008}. A suitable preparation protocol for quantum dark solitons is thus  to prepare a dark soliton in the small $\gamma$ regime, e.g.~by standard phase imprinting \cite{Burger1999,Denschlag2000}, possibly enhanced by density engineering \cite{Carr2001}, and then ramp the coupling strength $\gamma$ to the desired value by means of a Feshbach or confinement-induced resonance \cite{Haller2009}.

We have prepared the candidate quantum dark soliton of Eq.\ \eqref{eq:qds} as a Gaussian superposition of yrast states, and the properties of Gaussian wave packets have led us to hypothesise the equations for the width of the density feature (\ref{eq:varsum}) -- (\ref{eq:initvar}).
Given that these equations  are well supported by numerical evidence, we may hope that they can eventually be proven within the framework of the Bethe ansatz, and validated by experiments. 
While the Gaussian profile  of Eq.\ \eqref{eq:Gauss} was a somewhat arbitrary choice, it seems reasonable to expect that Eqs.\  (\ref{eq:varsum}) -- (\ref{eq:initvar}) are only true for Gaussian profiles, and that an uncertainty relation of the form
\begin{align}\label{eq:uncertainty}
\Delta P\sqrt{\Dx^2 - \sfs^2} \leq \frac{\hbar}{2} ,
\end{align}
holds for arbitrary superpositions in analogy to the well-known position--momentum uncertainty for point particles. In this more general context, $\Dx$ and $\Delta P$ represent measurable quantities while $\sfs$ is an intrinsic property of the dominant yrast state.
The Gaussian profile 
at $t=0$ then realises equality in the relation \eqref{eq:uncertainty} as a minimum uncertainty wave packet.
The Gaussian superposition thus presents an ``optimal quantum dark soliton'' by obeying  Eqs.\  (\ref{eq:varsum}) -- (\ref{eq:initvar}).
The properties of quantum states constructed using Bogoliubov theory in Ref.\ \cite{Dziarmaga2004} correspond to optimal quantum dark solitons in this sense, while the equal-weight superposition  of all yrast states in the interval $q\in [0, 2\pi n_0 \hbar)$ of Ref.\ \cite{Sato2016} falls outside of this framework.

The significance of the results presented here goes beyond the specific exactly-solvable model. The emerging picture of quasiparticle dynamics of yrast excitations in a strongly correlated quantum fluid is so simple and intuitive that we may expect it to be valid for non-integrable systems as well, e.g.~ultracold atoms with dipolar interactions, electrons in quantum wires, or Josephson vortices in coupled Bose gases \cite{Shamailov2018}. By simple extension, our framework allows for the study of soliton collisions, the results of which are left for a future publication.

\begin{acknowledgments}
We thank G.\ Astrakharchik for discussion. JB thanks the Max Planck Institute for Solid State Research for hospitality during a stay where part of this work was completed. This work was partially supported by the Marsden fund of New Zealand (contract number MAU1604) and by a grant from the Simmons Foundation. This work was performed in part at Aspen Center for Physics, which is supported by National Science Foundation grant PHY-1607611. SS was supported by the Massey University Doctoral Research Dissemination Grant.
\end{acknowledgments}

\bibliography{Sophie-save,Solitons}

\end{document}